\documentclass[twoside,fleqn]{article}

\usepackage{espcrc2}
\usepackage{rotating}
\usepackage{subfigure}
\usepackage{amssymb}
\usepackage{latexsym}
\usepackage{epic}
\usepackage{graphics}
\usepackage{epsfig}
\usepackage[bf,footnotesize]{caption}
\def\lsi{\raise0.3ex\hbox{$<$\kern-0.75em\raise-1.1ex\hbox{$\sim$}}}
\def\gsi{\raise0.3ex\hbox{$>$\kern-0.75em\raise-1.1ex\hbox{$\sim$}}}

\newcommand{\R}{{\kern+.25em\sf{R}\kern-.78em\sf{I} 
  \kern+.78em\kern-.25em}}
\newcommand{\C}{{\kern+.25em\sf{C}\kern-.50em\sf{I} \kern+.50em\kern-.25em}}

\title{Simulating non-commutative field theory
\thanks{Based on talks presented by W.B.\ and F.H.\ at Lattice02. 
\newline \hspace*{0.5mm} HU-EP-02/35.}}

\author{W. Bietenholz
\address{\vspace{-.25cm}Humboldt Universit\"at zu Berlin, 
    Invalidenstr. 110, D--10115 Berlin, Germany},
  F. Hofheinz$^{\mbox{\scriptsize a,}}
$\address{\vspace{-.25cm}Freie Universit\"at Berlin, 
    Arnimallee 14, D--14195 Berlin, Germany}
  and J. Nishimura
\address{Dept. Physics, Nagoya University, Nagoya 464--8602, Japan}}

\begin{document}

\begin{abstract}

Non-commutative (NC) field theories can be mapped onto
twisted matrix models. 
This mapping enables their Monte Carlo simulation, 
where the large $N$ limit of the matrix models describes 
the continuum limit of NC field theory. 
First we present numeric results
for 2d NC gauge theory of rank 1, which turns out to be 
renormalizable. The area law for the
Wilson loop holds at small area, but at large area we observe
a rotating phase, which corresponds to an Aharonov-Bohm 
effect. Next we investigate the NC $\phi^{4}$ model in $d=3$ and
explore its phase diagram.
Our results agree with 
a conjecture by Gubser and Sondhi in $d=4$, who predicted
that the ordered regime splits into a uniform
phase and a phase dominated by stripe patterns.

\vspace*{-5mm}

\end{abstract}

\maketitle

\section{INTRODUCTION}

In the recent years there has been tremendous activity in the research of
field theories on NC spaces. One field of application of such theories
is the quantum Hall effect \cite{condmat}. The boom of interest, however,
was triggered by the observation that string and M theory at low energy 
correspond to NC field theories, see e.g.\ Ref.\ \cite{SW99}.

The NC space is characterized by the non-commutativity of some of its
coordinates. Here we are concerned with the case where only two
coordinates do not commute,
\begin{equation}  \label{NCspace}
[\hat x_{\mu}, \hat x_{\nu} ] = i \theta \epsilon_{\mu \nu} \ .
\end{equation}
For a review of field theory on NC spaces (``NC field theory''), 
see Ref.\ \cite{revNC}. The non-commutativity yields non-locality
in a range of $O(\sqrt{\theta})$. This property implies conceptual 
problems, but there is also hope that it might provide the crucial
link to string theory or quantum gravity. From the technical point
of view, it makes the perturbative renormalization even harder
than in commutative field theory, because a new type of singularity
occurs, which is UV and IR at the same time.
Heuristically this can be understood immediately from relation
(\ref{NCspace}) along with Heisenberg's uncertainty relation.
So far, there is no systematic
machine for absorbing this kind of mixed UV/IR divergences.

\section{GAUGE THEORY ON A NC PLANE}

In particular, the NC rank 1 gauge action can be written in a form 
which looks similar to the familiar (commutative) case,
\begin{eqnarray}
&& \hspace*{-5mm}
S [A] = \frac{1}{4} \int d^{d}x \, {\rm Tr} (F_{\mu \nu}
\star F_{\mu \nu}) \\
&& \hspace*{-5mm}
F_{\mu \nu} = \partial_{\mu} A_{\nu} - \partial_{\nu} A_{\mu}
+ ig (A_{\mu} \star A_{\nu} - A_{\nu} \star A_{\mu}) \ , \nonumber
\end{eqnarray}
where the fields are multiplied by the star product
\begin{equation}
f(x) \star g(x) = e^{\frac{1}{2} i \theta \epsilon_{\mu \nu}
\frac{\partial}{\partial x_{\mu}} \frac{\partial}{\partial y_{\nu}}} 
f(x) g(y) \vert_{x=y} .
\end{equation}
This action is star gauge invariant. Also lattice formulations
exist, but they can hardly be simulated: one would need star 
unitary link variables, construct a measure for them etc.

In 1999 Ishibashi et al.\ observed \cite{IIKK} that NC $U(n)$ gauge 
theories can be mapped onto specific versions of the twisted 
Eguchi-Kawai model (TEK) \cite{TEK}, which has the action
\begin{eqnarray}
&& \hspace*{-7mm}
S_{TEK}[U] = - N \beta \sum_{\mu \neq \nu} Z_{\mu \nu}
\, {\rm Tr} (U_{\mu} U_{\nu} U_{\mu}^{\dagger} U_{\nu}^{\dagger}) \ , \\
&& \hspace*{-7mm}
Z_{\mu \nu} = e^{ 2 \pi i k/L } = Z_{\nu \mu}^{*} \ \
(\mu \! < \! \nu , \, k \in \mathbb{Z} , \, L = N^{2/d} ) \, . \nonumber
\end{eqnarray}
Since this system lives on one point there is only one $N\times N$
link variable in each direction. The factor $Z_{\mu \nu}$ is the twist.
This mapping is based on Morita equivalence
(an exact equivalence of the corresponding algebras)
in the large $N$ limit
\cite{IIKK}, and on discrete Morita equivalence at finite $N$, which
corresponds to NC gauge theory on the lattice \cite{AMNS}.
The equivalence holds if we choose 
\begin{equation}  \label{twist}
k = \frac{N+1}{2} \ , \quad {\rm and~it~implies} \quad 
\theta =\frac{1}{\pi} N a^{2} \ ,
\end{equation}
where $N$ must be odd, and
$a$ is the lattice spacing in the NC space.
Also the Wilson loops, which are defined in the TEK in the obvious way
(see below) are mapped onto the Wilson loops in NC gauge theory.

This provides a non-perturbative access to NC gauge theory.
The problems with the direct NC lattice formulation are
all circumvented by this mapping onto a matrix model;
hence this is a real counter example to the ``no-free-lunch
theorem''.\\

Here we are interested in pure NC $U(1)$ gauge theory
in $d=2$ and we summarize our numeric results of
Ref.\ \cite{2dU1}. For analytical work on this model,
see in particular Ref.\ \cite{PanSza}.

Due to Eguchi-Kawai equivalence, the large $N$ limit of the TEK
at fixed $\beta$ coincides with the planar large $N$ limit of
commutative gauge theory, which has been solved analytically 
\cite{GW}. In that case the Wilson loop obeys an exact area law,
\begin{eqnarray}  \label{arealaw}
\langle W(I\times J ) \rangle &=& e^{-\kappa(\beta ) IJ} \\
\kappa (\beta ) &=& - \ln (1 - \frac{1}{4 \beta }) \quad {\rm at } \quad
\beta \geq \frac{1}{2} \ . \nonumber
\end{eqnarray}
From that framework we adapt the ``physical area'' $a^{2}IJ$,
where $a = \sqrt{\kappa}$. However, for our purposes it is crucial to
take the {\em double scaling limit},
which keeps $N a^{2} \simeq N /\beta $ constant, so that
the corresponding NC parameter $\theta$ remains finite
in the large $N$ limit, c.f.\ eq.\ (\ref{twist}).

This is different from the planar large $N$ limit; the latter
would lead to the Gross-Witten solution, in agreement
with the well-known equivalence of $\theta \to \infty$
and commutative gauge theory in the planar large $N$ limit.

The 2d TEK was simulated before \cite{TEK2d}, but not with
the parameters required here.
Thanks to a trick introduced in Ref.\
\cite{heatbath} we could apply a heat bath algorithm.

\begin{figure}[htbp]
 \begin{center} 
   \vspace{-5mm}
   \includegraphics[width=0.99\linewidth]{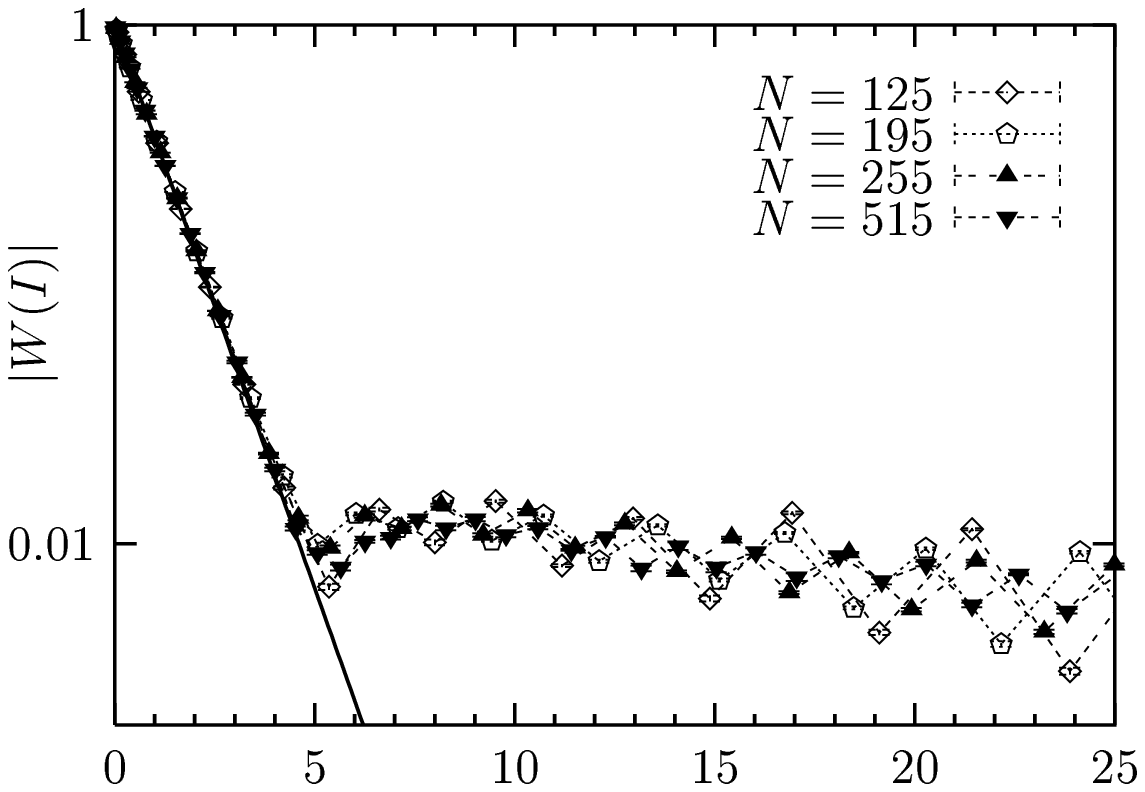}

   \vspace{0.2cm}
   \hspace{.2cm}\includegraphics[width=.95\linewidth]{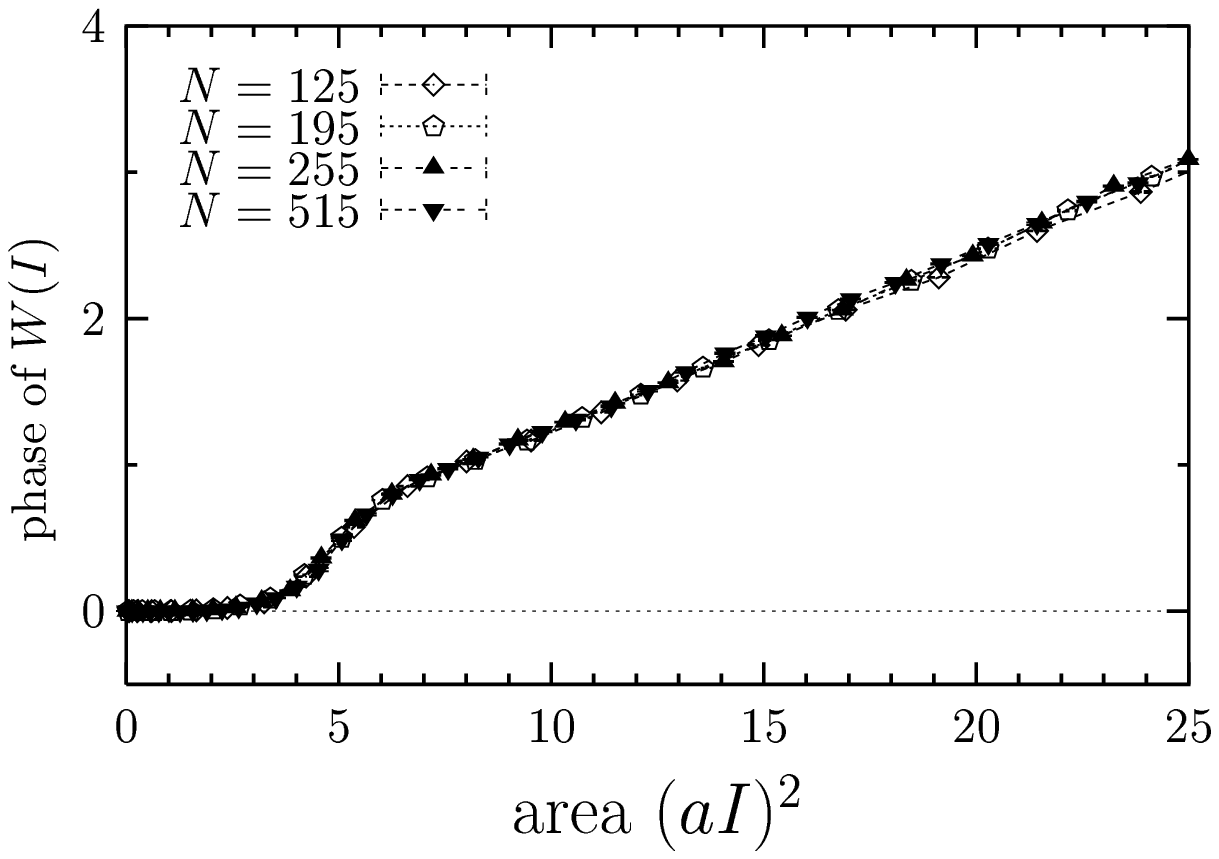}
 \end{center}
\vspace{-10mm}
 \caption{{\it The polar coordinates of the complex Wilson loop $W(I)$
 plotted against the physical area $A = a^2 I^2$.
 At small areas it follows the Gross-Witten
 area law (solid line). 
 At larger areas the absolute value does not decay any more, 
 but the phase increases linearly.}}
\label{wil_polar}
\vspace*{-5mm}
\end{figure}

We focus on the square-shaped Wilson loop,
\begin{eqnarray}
W_{12}(I\times I) &=& \frac{1}{N} Z_{12}^{I^{2}} \, {\rm Tr}
(U_{\mu}^{I} U_{\nu}^{I}U_{\mu}^{\dagger \, I} U_{\nu}^{\dagger \, I}) 
\nonumber \\
&=& W_{21}^{*} (I \times I) \ .
\end{eqnarray}
Note that $W(I) := \langle W_{12}(I\times I )\rangle$ is complex because of
the twist. In Fig.\ \ref{wil_polar} we show its behavior in 
polar coordinates at $N/\beta =32$. 
We see that a double scaling limit does manifestly exist.
Since it corresponds to the continuum limit of 2d NC rank 1 gauge theory,
we conclude that its Wilson loop is non-perturbatively renormalizable.
At small area (with respect to $O(\theta )$) it is 
practically real and follows the Gross-Witten area law of eq.\ 
(\ref{arealaw}).
Beyond that regime $\vert W(I)\vert$ does not decay
any more, but the phase grows {\em linearly} in the area.

\begin{figure}[htbp]
 \begin{center}
   \vspace*{-3mm}
   \includegraphics[width=.92\linewidth]{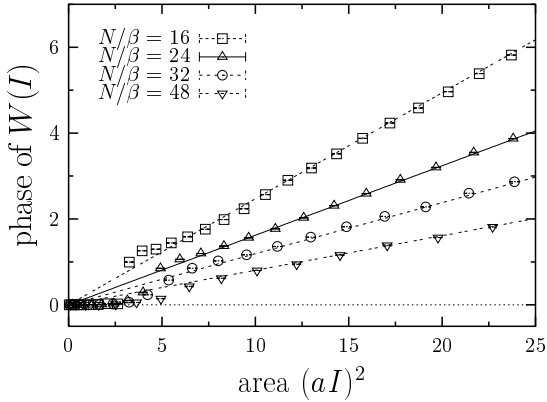}
 \end{center}
\vspace{-10mm}
 \caption{{\it The phase of the Wilson loop $W(I)$ at $N=125$ for various
values of $N/\beta$, corresponding to different values of $\theta$.
At large areas $A >  O(\theta )$ the phase $\phi$ agrees with 
the formula $\phi = A/\theta$, which is shown by straight lines.}} 
   \label{wil_phase}
   \vspace*{-5mm}
\end{figure}

Fig.\ \ref{wil_phase} shows the asymptotic linear behavior of the phase 
$\phi$ at different values of $N/\beta \simeq \pi \theta$. 
In the large area regime, we observe the simple relation
\begin{equation}
\phi = A / \theta \ , \qquad (A = (aI)^{2} = {\rm physical~area})
\end{equation}
to a very high accuracy. We checked that it holds more
generally for rectangular Wilson loops. Hence we can identify this
law with the Aharonov-Bohm effect, if we formally introduce a
constant magnetic field
\begin{equation}
B = 1 / \theta
\end{equation}
across the plane. In fact, this corresponds exactly to the relation
used by Seiberg and Witten when they mapped open strings in a
constant gauge background onto NC gauge theory \cite{SW99}.
Moreover, the very same relation was used in the description of the
quantum Hall effect, where an electron in a layer was projected to the 
lowest Landau level \cite{condmat}.
Here we recover this law as a dynamical effect.

We fix again $N/\beta =32$, and
we consider now the connected Wilson 2-point function
\begin{equation}
G_{2}^{(W)} = 
\langle W_{\mu \nu}(I \times I) W_{\nu \mu}(I \times I) \rangle_{c}
\in \mathbb{R} \ .
\end{equation}
After a wave function
renormalization $G_{2}^{(W)} \to \beta^{-0.6} G_{2}^{(W)} $
we observe also here a double scaling regime, see Fig.\ 3 (on top).

Finally we consider the Polyakov lines
\begin{equation}
P_{\mu}(I) = {\rm Tr} (U_{\mu}^{I}) \, , \
P_{-\mu}(I) = {\rm Tr} (U_{\mu}^{\dagger \, I}) \ .
\end{equation}
Note that $\langle P_{\mu}(I) \rangle =0$ due to the phase symmetry
(which corresponds to translation invariance), but the Polyakov 
multi-point functions are sensible observables.
Fig.\ 3 (below) shows that also the Polyakov 2-point function
\begin{equation}
G_{2}^{(P)}(I) = \langle P_{\mu}(I) P_{-\mu}(I) \rangle
\end{equation}
has a double scaling regime, if we apply the same wave function
renormalization again,
$G_{2}^{(P)} \to \beta^{-0.6} G_{2}^{(P)} $.
\begin{figure}[htbp]
 \begin{center} 
   \includegraphics[width=.95\linewidth]{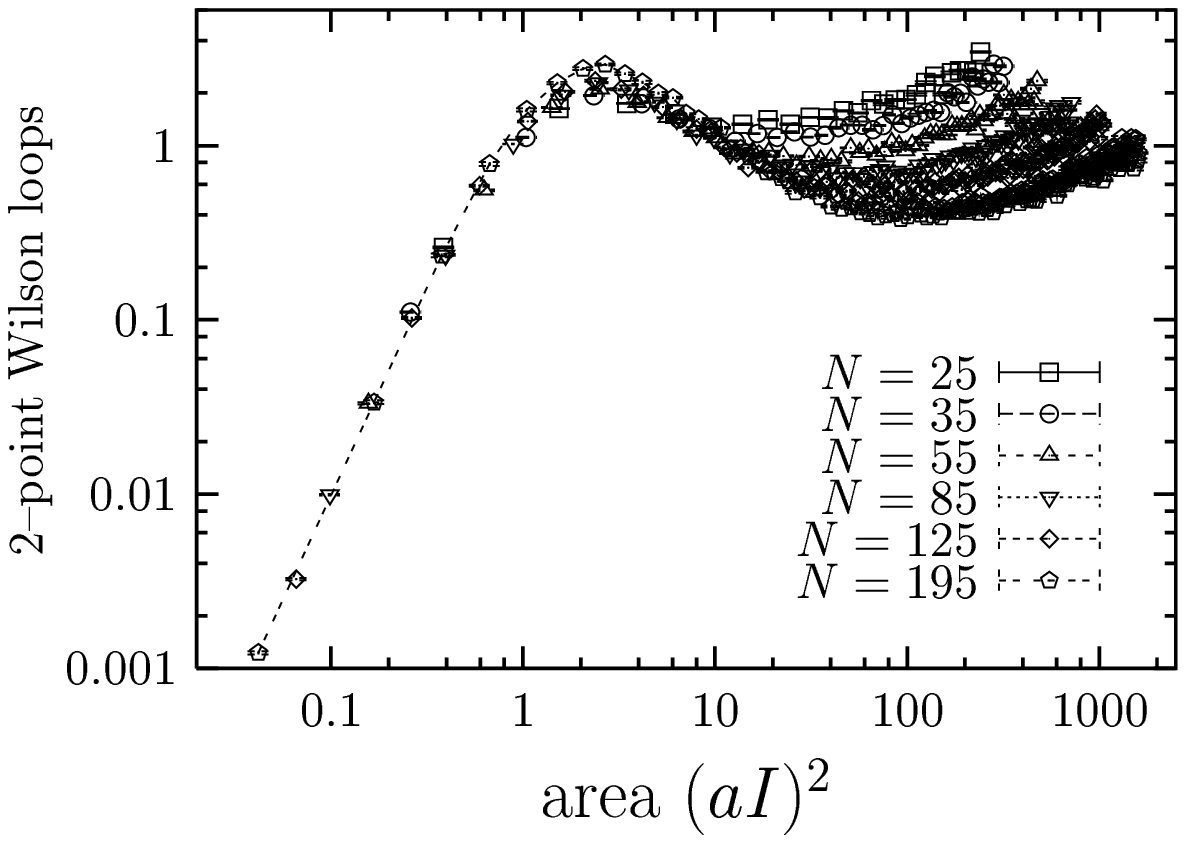} 

   \vspace*{8mm}
   \hspace*{.4cm}\includegraphics[width=.95\linewidth]{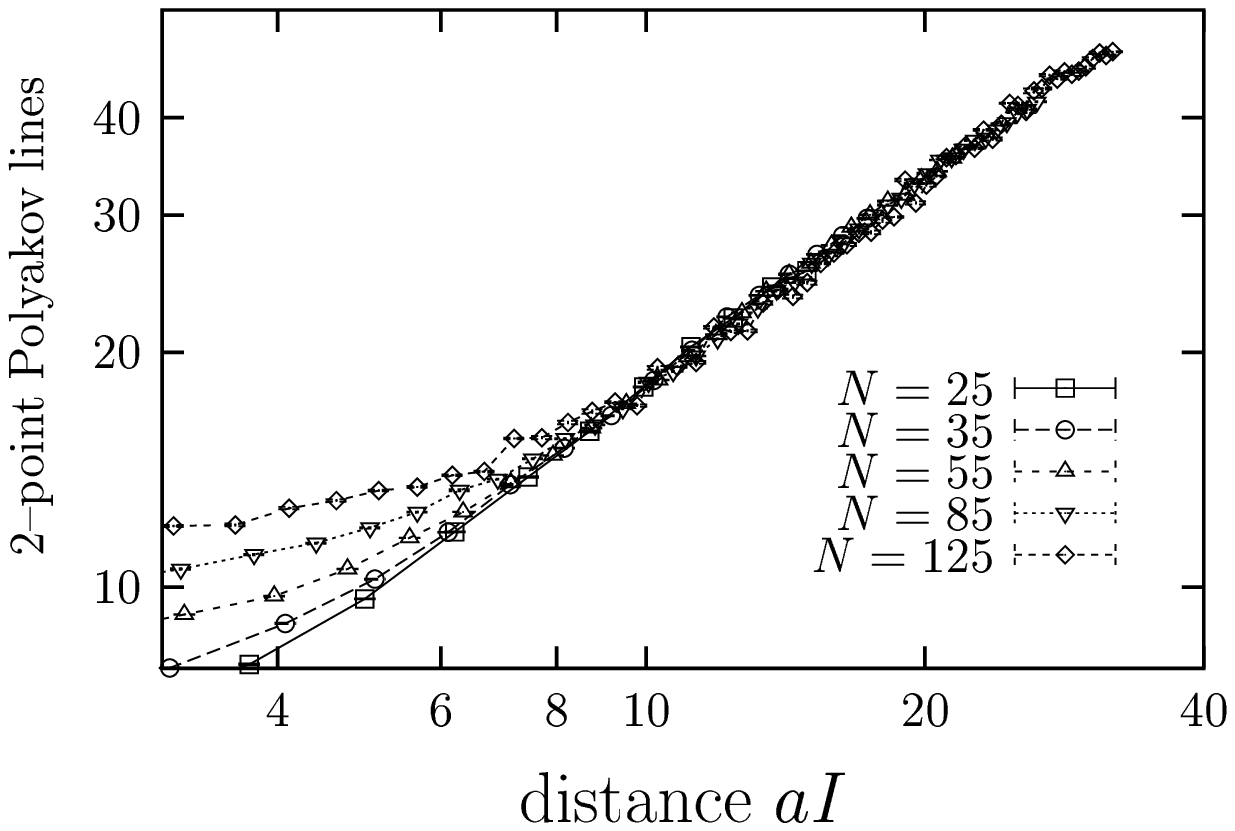}
 \end{center}
\vspace{-10mm}
 \caption{{\it The Wilson 2-point function $G_{2}^{(W)}$ (on top)
and the Polyakov 2-point function $G_{2}^{(P)}$ (below),
at $N=195$. After performing the same wave function renormalization
we find a double scaling regime for each of the 2-point functions.}}
\label{2point}
\vspace*{-8mm}
\end{figure}

\section{THE NC $\phi^{4}$ MODEL IN $d=3$}

We now summarize our ongoing work on the NC $\phi^{4}$ model in $d=3$
\cite{prep}. We investigate the phase diagram and compare our
observations in particular with a conjecture by Gubser and Sondhi,
which was obtained in $d=4$ by a 
Hartree--type
approximation \cite{GS}. They studied the phase--diagram in the
$m^2/\Lambda^2$--$\theta\Lambda^2$ plane for fixed coupling $\lambda$,
where $\Lambda$ is a momentum cut--off. 
The conjecture can be summarized as follows:
\begin{itemize}
\item For small $\theta$ there is an Ising type phase transition
  between a disordered and {\em uniformly} ordered phase.
\item For large enough $\theta$ this transition changes its nature, due to
  UV/IR effects. This phase transition is driven by a mode of the scalar
  field with non--vanishing momentum, leading to a non--uniformly ordered
  or {\em striped phase}. This phase transition is first order in the 
  regularized theory.
\item 
  The theory is renormalizable within the one--loop self--consistent 
Hartree--type
  approximation and the first order phase transition turns into second
  order in the continuum limit.
\end{itemize}
The action of the NC $\lambda\phi^4$ theory reads
\begin{equation}
  \label{eq:phi-action}
  S=\int d^dx\left[\frac{1}{2}\partial_\mu \phi\,\partial_\mu 
\phi+\frac{m^2}{2}\phi^2+\frac{\lambda}{4}\phi^{\star 4}\right]\,,
\end{equation}
where the interaction term is a star--product of four scalar fields. We study
this model in $d\!=\!3$, and we exclude the time
direction from non--commutativity. The spatial coordinates satisfy
eq.\  (\ref{NCspace}).

A lattice version of this action can be constructed, but as in Section 2
we map the system on a dimensionally reduced model to enable
Monte--Carlo simulations. This mapping was already carried out in 
Ref.\ \cite{AMNS}. In our case
the scalar field $\phi(\vec{x},t)$
defined on a $N^2T$ lattice is mapped on $N\times N$ Hermitian
matrices $\hat{\phi}(t)$. Their action takes the form
\begin{eqnarray}
  \label{eq:reduced-action}
      &\hspace{-1.4cm}S=\textrm{Tr}\sum_{t=1}^{T} \biggl[
      \frac{1}{2}\sum_\mu\left(\Gamma_\mu\hat{\phi}(t)\,
\Gamma_\mu^\dagger-\hat{\phi}(t)\right)^2\\
  &\hspace{.2cm}+\frac{1}{2}\left(\hat{\phi}(t+1)-\hat{\phi}(t)\right)^2+
  \frac{m^2}{2}\hat{\phi}^2(t)+\frac{\lambda}{4}\hat{\phi}^4(t)
\biggl]\nonumber\,.
\end{eqnarray}
There are two kinetic terms according to the two underlying
geometries. The matrices $\Gamma_\mu$ are called   'twist eaters'  and act
here as shift operators. They are defined by the relation
\begin{equation}
  \label{eq:twisteaters}
  \Gamma_\mu\Gamma_\nu=Z^*_{\mu\nu}\Gamma_\nu\Gamma_\mu\,,
\end{equation}
where $Z_{\mu\nu}$ is the twist already introduced in Section 2.
Again we have to use odd values of $N$.

For this action we study the phase diagram, but in contrast to
Gubser and Sondhi we consider the $m^2$--$\lambda$
plane at fixed $\theta$. By analogy we expect a striped phase for
large enough $\lambda$. 

\begin{figure}[htbp]
  \centering
  \includegraphics[width=.95\linewidth]{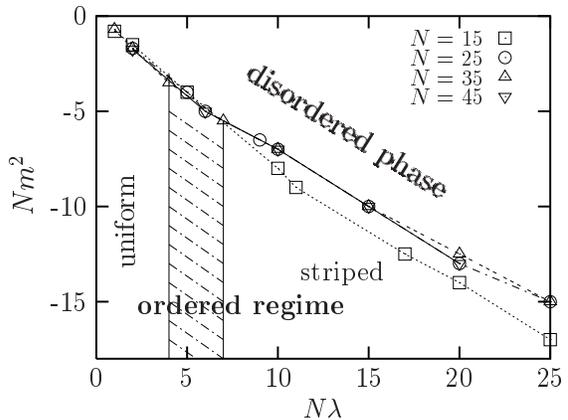}\\
  \vspace{-1cm}
  \caption{\it The phase diagram of the 3d NC $\phi^4$ theory. The connected points 
    show the separation line between disordered and ordered regime,
    and the hatched region marks the transition between uniformly
    ordered and striped phase at $N=35$.}
  \label{fig:phasediagram}
\end{figure}

Our results for $N=15\dots 45$ are summarized in the phase
diagram in Fig.\ \ref{fig:phasediagram}.  \vspace*{-.5cm}

\vspace*{-.7cm} So far we have been able to identify the separation
line between ordered regime and disordered phase and we observe a
large $N$ scaling. This separation is shown by the points
connected with lines.  The vertical lines indicate the transition
region between uniformly ordered and striped phase at $N=35$. For
$N<35$ we find the transition between the two ordered regimes at
larger values of $N\lambda$. Below we discuss the
procedure that led us to this phase diagram.\\

\begin{figure}[htbp]
  \centering
  \hspace*{-.1cm}\includegraphics[width=.95\linewidth]{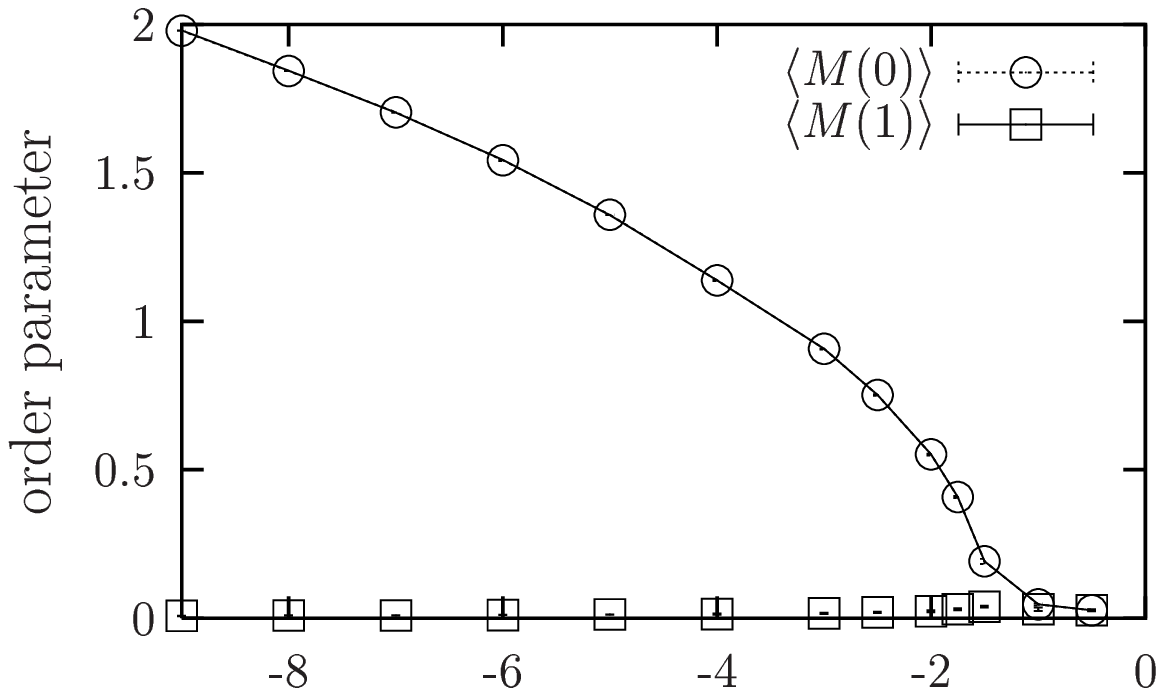}

  \vspace*{.4cm}
  \includegraphics[width=.95\linewidth]{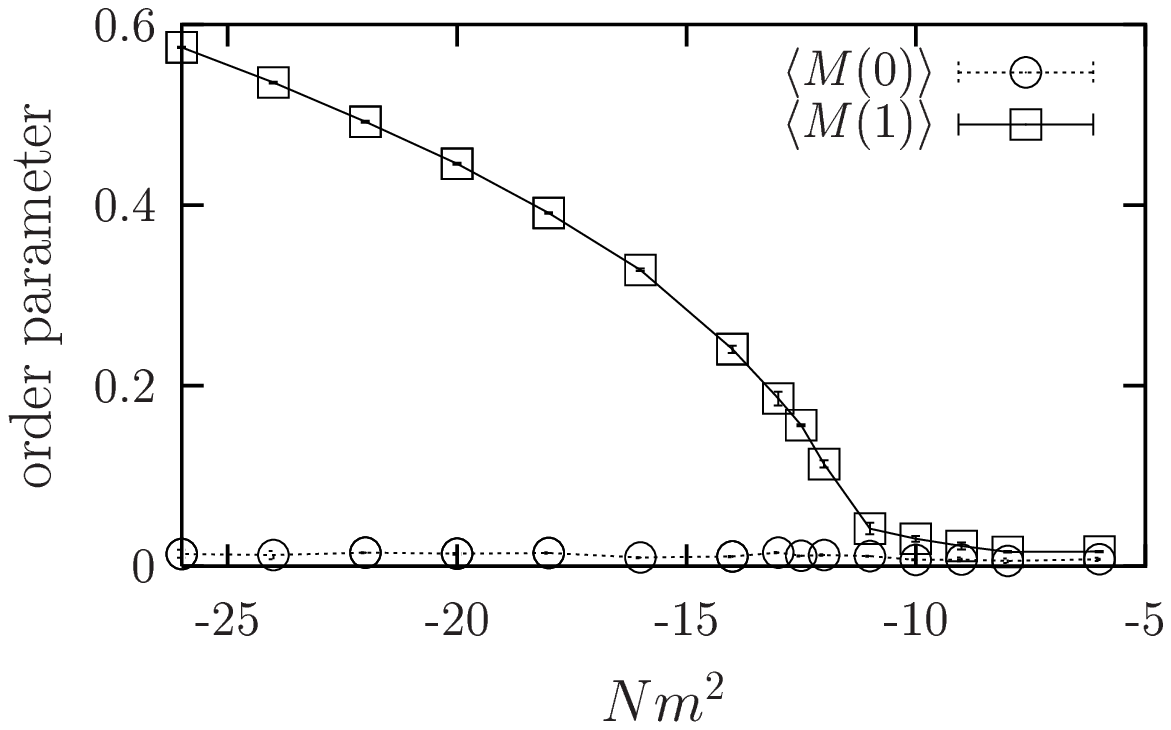}
  \vspace{-1cm}
  \caption{\it The momentum dependent order parameter against $Nm^2$.
    Above we fixed $N\lambda=2$, which leads to a
    uniformly ordered phase, and below at $N\lambda=20$ leading
    to a striped phase.}
  \label{fig:phase1}
\end{figure}

We use the momentum dependent order parame\-ter
$\langle M(k) \rangle$, where
\begin{equation}
  \label{eq:orderparameter1}
  M(k):=\frac{1}{NT} \, \max_{|\vec{p}|=k} \,
  \bigl| \, \sum_t\tilde{\phi} (\vec{p},t) \, \bigl| \ ,
\end{equation}
and $\tilde{\phi} (\vec{p},t)$ is the spatial Fourier transform
of $\phi (\vec{x},t)$ \
\footnote{Note that $\tilde{\phi} (\vec{p},t)$ 
coincides with the Fourier transform
in the commutative case.}.
In particular $\langle M(0)\rangle $ reduces to the standard 
order parameter for
the spontaneous breakdown of a $Z_2$ symmetry. To find the separation
line between ordered and disordered phase, we  decreased $m^2$ at constant
values of $\lambda$, starting in the disordered phase. 
Two examples at $N=35$ are shown in Fig.\ \ref{fig:phase1} 
\vspace*{-.5cm}

\vspace*{-.7cm} In these plots we show the standard order parameter
$\langle M(0) \rangle$ and the staggered order parameter 
$\langle M(1) \rangle$.
In the upper plot of Fig.\ \ref{fig:phase1} only $\langle M(0)\rangle$
is non--trivial below some value of $m^2<0$. 
Therefore we are in the uniformly ordered phase. In the lower
plot of Fig.\ \ref{fig:phase1}, $\langle M(0)\rangle $ practically
vanishes for all $m^2$, but $\langle M(1) \rangle $ is
non--zero, indicating a striped phase.  Such measurements allow us to
identify the separation region between uniformly ordered and striped
phase.

\begin{figure}[htbp]
  \centering
  \hspace*{-.1cm}\includegraphics[width=.95\linewidth]{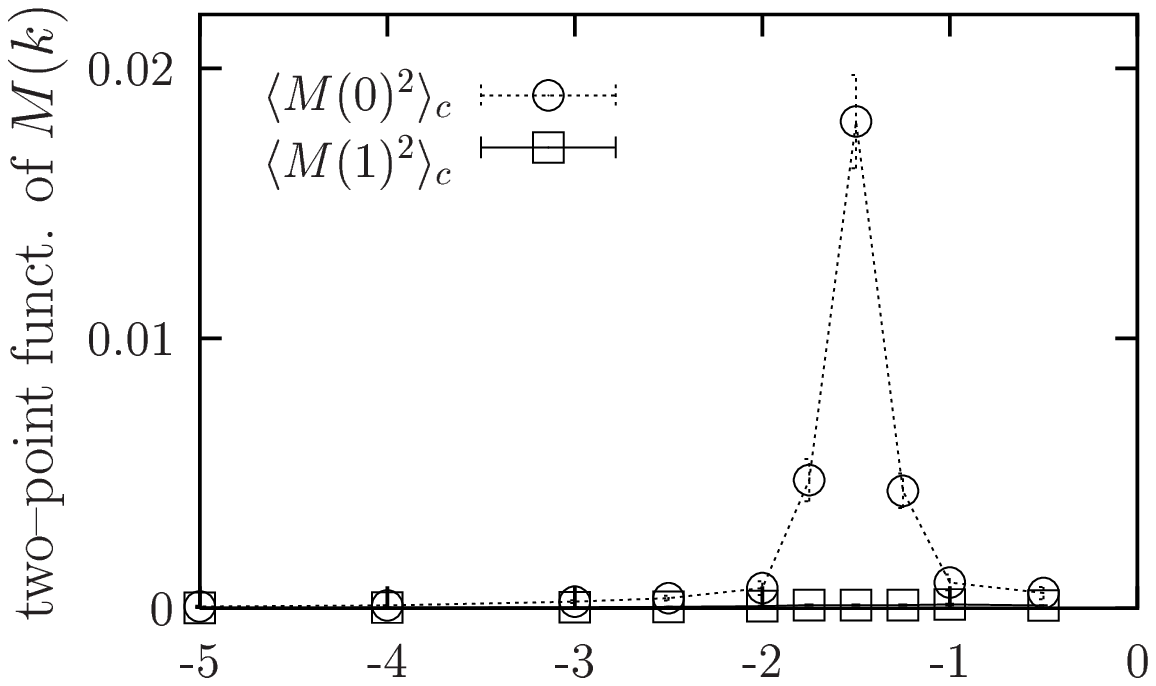}

  \vspace*{.4cm}
  \includegraphics[width=.95\linewidth]{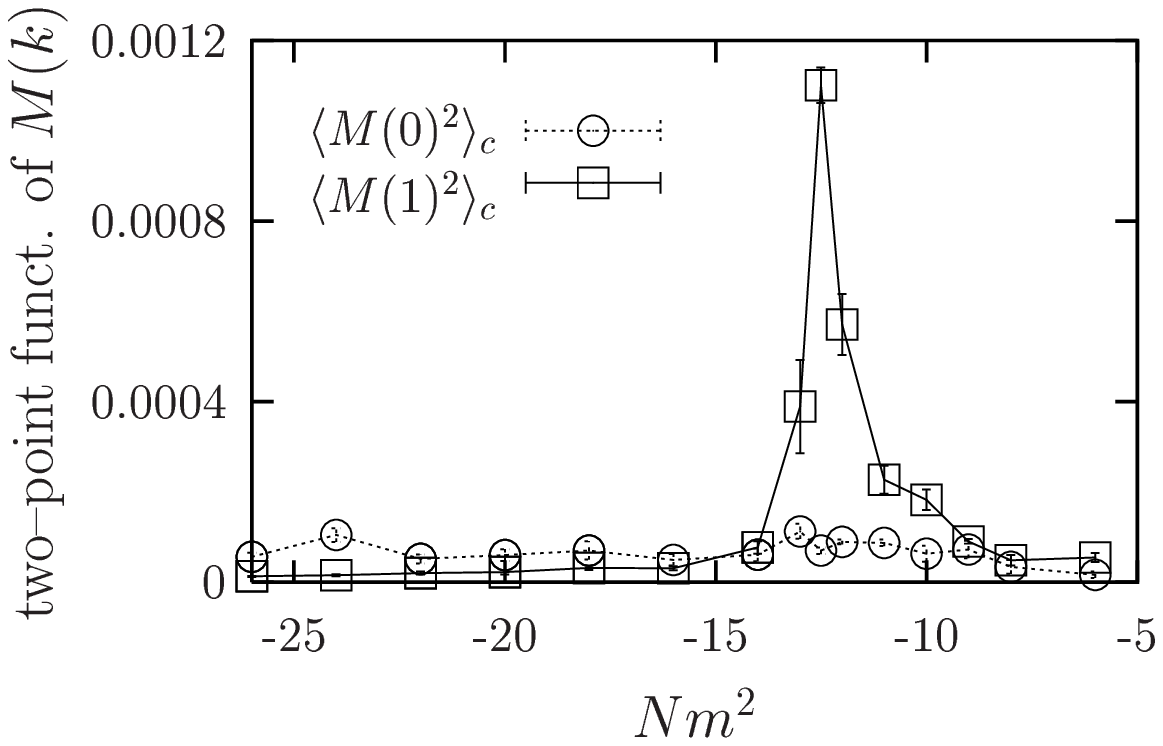}
  \vspace{-1cm}
  \caption{\it The two--point function of the quantity
    $M(k)$, defined in eq.\ (\ref{eq:orderparameter1}),
    against $Nm^2$ at $N\lambda=2$ (above)
  and at $N\lambda=20$ (below).}
  \label{fig:phase2}
\end{figure}

To localize the separation line between ordered and disordered phase
we computed the connected part of the two--point function of $M(k)$.
This function has a peak at the phase transition.
Fig.\ \ref{fig:phase2} illustrates some examples for these measurements.

\begin{figure}[htbp]
  \centering
  \subfigure[{$Nm^2=-0.5$}]{\epsfig{figure=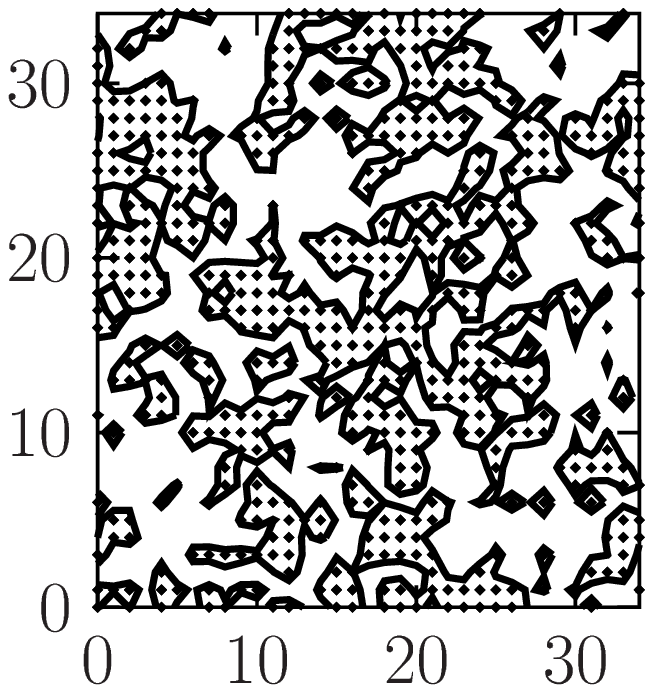,width=.41\linewidth}}%
  \hspace*{1cm}\subfigure[{$Nm^2=-8$}]{\epsfig{figure=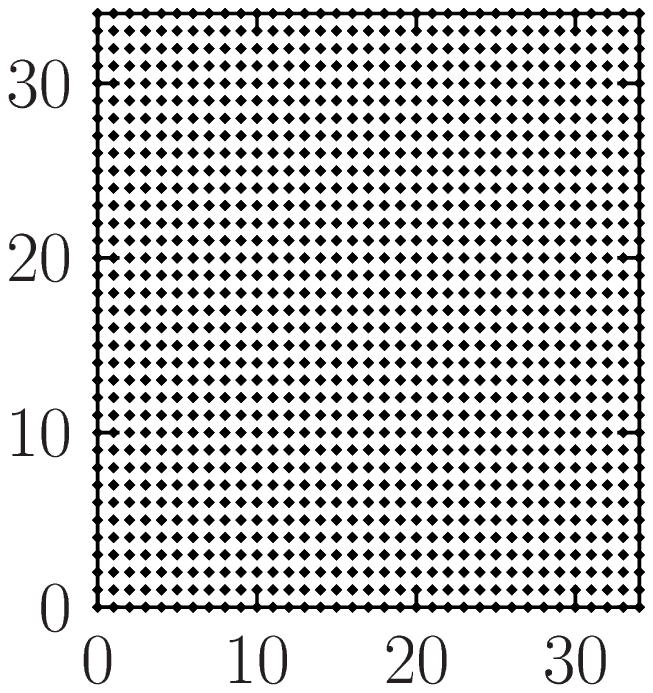,
width=.41\linewidth}}\\

  \vspace{-1cm}
  \begin{center}
    $N\lambda=2$
  \end{center}

  \vspace{.2cm}
  \subfigure[$Nm^2=-6$]{\epsfig{figure=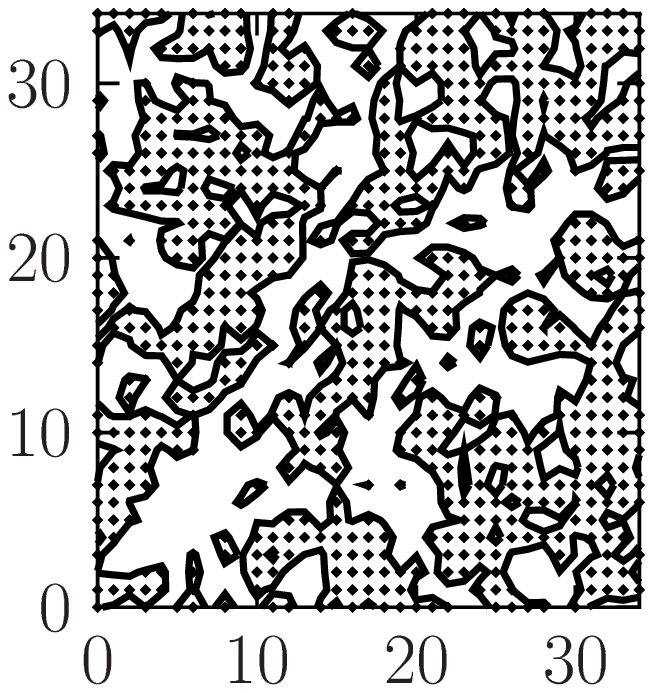,width=.415\linewidth}}%
  \hspace*{1cm}\subfigure[$Nm^2=-26$]
{\epsfig{figure=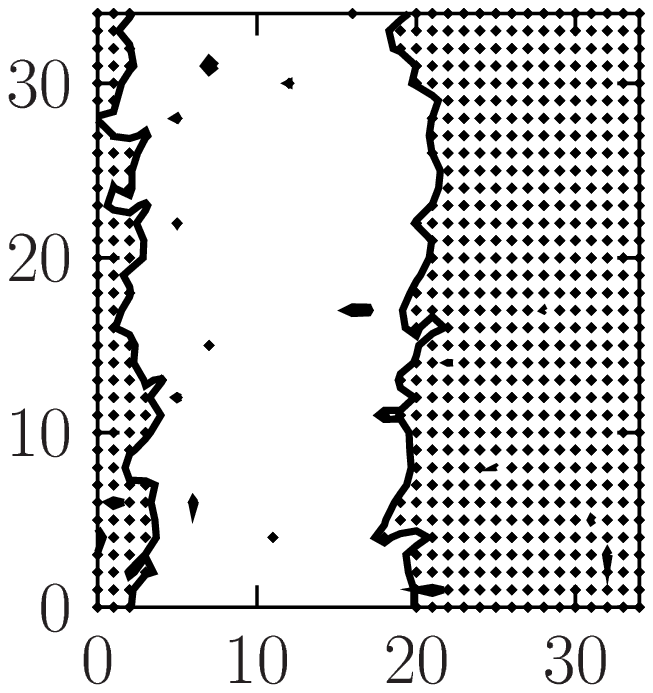,width=.415\linewidth}}\\

  \vspace{-1.cm}
  \begin{center}
    $N\lambda=20$
  \end{center}

  \vspace{-1cm}
  \caption{\it Snapshots of single configurations 
$\phi(\vec{x},t)$ at a certain time $t$, for $N=35$.}
  \label{fig:stripe}
\end{figure}

\vspace*{-.8cm} 
In Fig.\ \ref{fig:stripe} we show some snapshots of
configurations $\phi(\vec{x},t)$ at $N=35$. 
The dotted areas mark the regions where $\phi>0$, and the blank areas
represent $\phi<0$. At $N\lambda=2$ and for small $|Nm^2|$ (Fig.\ 
\ref{fig:stripe}a) we find such areas scattered all over the volume, and
at $Nm^2\ll0$ (Fig.\ \ref{fig:stripe}b) $\phi(\vec{x},t)$ is either
positive or negative for all $x$, as one expects for an Ising type
phase. At $N\lambda=20$ we also find a disordered phase (Fig.\ 
\ref{fig:stripe}c), but for $Nm^2\ll0$ the uniformly ordered phase
is replaced by a striped phase (Fig.\ \ref{fig:stripe}d). 
\vspace*{-.4cm} 

\vspace*{-1.1cm} 
\section{CONCLUSIONS}

We presented numeric results for NC field theory, which could
be obtained through the mapping onto twisted matrix models.

In the first part we summarized the results of Ref.\ \cite{2dU1}.
Specific types of the TEK correspond to 2d NC $U(1)$ gauge
theory on the lattice. We approached the large $N$ limit so that the 
non-commutativity parameter $\theta$ is kept constant (double scaling 
limit). This describes the continuum limit of NC gauge theory.
We found finite limits for the 1- and 2-point Wilson loop, as well
as the 2-point Polyakov line, which confirms that this model is
non-perturbatively renormalizable. This is the first evidence for the
non-perturbative renormalizability of a NC field theory. 

At small areas, the Wilson loop follows an area law
as in the (commutative) planar theory.
At larger areas it deviates and reveals a new
double scaling limit. This is a manifestation of non-perturbative
UV/IR mixing. In that regime, the phase of the Wilson
loop is given simply by the area divided by $\theta$. This corresponds
to an Aharonov-Bohm effect if we identify $\theta$ with a magnetic field
$B = 1/\theta$, as it is also done in the solid state \cite{condmat} and 
string \cite{SW99} literature.\\

\vspace{-.3cm}
We further studied the phase diagram of a 3d NC $\phi^4$ theory \cite{prep}. 
The separation line between ordered and disordered phase is localized 
and we observe large $N$ scaling. In agreement with the qualitative 
conjecture in Ref.\ \cite{GS},
the ordered regime splits into a uniform and a striped phase. \\

{\small \noindent {\bf Acknowledgment:} 
We thank A. Barresi, H.\ Dorn, Ph.\ de Forcrand 
and R.\ Szabo for useful comments.}

\end{document}